\documentclass[aps,pre,twocolumn,showpacs,groupedaddress]{revtex4-1}
\usepackage{graphicx}
\begin{document}

\title{Probability distribution of the order parameter in the directed percolation universality class}

\author{P. H. L. Martins}
\email[]{pmartins@fisica.ufmt.br}
\affiliation{Instituto de F\'{\i}sica, Universidade Federal de Mato Grosso, Av. Fernando Corr\^ea da Costa, 2367,  Cuiab\'a, MT, 78060-900, Brazil}
\affiliation{Center for Simulational Physics, University of Georgia, Athens, GA 30602, USA}

\date{\today}

\begin{abstract}
The probability distributions of the order parameter for two models in the directed percolation universality class were evaluated. 
Monte Carlo simulations have been performed for the one-dimensional generalized contact process and the Domany-Kinzel cellular 
automaton. In both cases, the density of active sites was chosen as the order parameter. The criticality of those models was 
obtained by solely using the corresponding probability distribution function. It has been shown that the present method, which has been successfully employed in treating equilibrium systems, is indeed also useful in the study of nonequilibrium phase transitions.
\end{abstract}

\pacs{05.50.+q, 05.70.Ln, 05.70.Jk, 05.10.Ln}

\maketitle

\section{\label{intro}Introduction}

The theoretical treatment of nonequilibrium dynamic systems has been given a great deal of attention during the last decades. 
However, even the simplest models have not yet reached the same level of understanding as their equilibrium counterparts. 
Excellent resources on this subject can be found in Refs. \cite{2,3,4} and references therein.
 
One of the most important models in nonequilibrium physics is directed percolation (DP), that 
is widely considered as an analogue of the Ising model for nonequilibrium phase transitions. 
DP has been used to simulate a large variety of problems, including flow of a liquid through a porous medium, 
electric current in a diluted diode network, and reaction-diffusion processes. 
Another interesting model is the so-called contact process (CP), which was first introduced by Harris\cite{1} as a nonequilibrium toy model 
to study epidemic spreading. In the standard CP, each site of a lattice can be active, representing an infected individual, 
or inactive, corresponding to a healthy person. The system evolves in such a way that only one site is updated at a time.
In a $d$-dimensional hyper-cubic lattice, the annihilation rate $\mu$ means that an active (occupied) site becomes inactive 
(vacant) at rate $\mu$, independent of its neighbors. A vacant site turns to occupied at a creation rate proportional to 
the fraction of occupied neighbors, $n/2d$, where $n$ is the number of occupied nearest neighbors. 
Thus, an inactive site surrounded only by inactive neighbors remains inactive. 
Once all sites are vacant, the system becomes trapped in that state, which is known as frozen or absorbing state.
It is well known that the CP undergoes a second order phase transition from an active to a frozen (absorbing) phase at some critical $\mu_c$. 
For annihilation rates $\mu > \mu_c$, the only quasi-stationary state is the absorbing one, while for sufficiently small $\mu$ 
a finite fraction of sites remains active. In spite of its simplicity, no exact results for $\mu_c$ are known, even for one-dimensional lattices. 
Mean field approximation, Monte Carlo simulations, and series expansion are the most used techniques\cite{5,6,7,8}. 
The best estimate for the critical point in one dimension\cite{7,8} is $\mu_c = 0.303228(2)$.

A generalized version of the CP is obtained by considering different creation rates. For instance, in one dimension, 
let us call $\zeta$ the creation rate at a given site with exactly one occupied neighbor. 
The creation rate with two occupied neighbors is set to $1$, while annihilation occurs at rate $\mu$. 
Standard CP corresponds to $\zeta = 0.5$, while $\zeta = 1.0$ is known as the $A$-model\cite{8b}. 
This family of processes was proposed by Durret and Griffeath\cite{9} and has received some attention later on\cite{8,10}. 
Analogous to the usual CP, this generalized contact process (GCP) also shows a continuous phase transition 
from the active state to the absorbing state for infinite systems. A different generalization of the contact process, 
although not discussed in the present study, considers more than one absorbing state\cite{10b} and 
is also a subject of current interest\cite{10c,10d}.

Another irreversible model that describes a nonequilibrium phase transition from an active to an absorbing state 
is the probabilistic Domany-Kinzel cellular automaton (DKCA)\cite{14}.
It is represented in a one-dimensional lattice containing $N$ sites that can be either empty or occupied. 
The state of each site $i$ at time $t+1$ depends only upon the state of the two nearest neighbors 
$i-1$ and $i+1$ at time $t$. In contrast to the GCP, in the DKCA all sites are updated simultaneously.  
Denoting the occupation variable by $\sigma_i$, one can define the conditional probabilities 
$P[\sigma_i(t+1)|\sigma_{i-1}(t),\sigma_{i+1}(t)]$, where $\sigma_i(t) = $1(0) if the site $i$ is 
occupied (empty) at time $t$. 
Besides the active/absorbing transition, this model exhibits a damage spreading transition in the active phase, 
from a chaotic to a non-chaotic region\cite{15,16}. 

Regarding universality classes, it is believed that most models having 
absorbing state transitions belong to the directed percolation universality class 
since some essential features like short-range interactions and translational invariance 
are fulfilled\cite{11,12,13}. Thus, the absorbing transitions in the GCP and the DKCA models are in the DP universality class\cite{17}.
From an experimentalist point of view, this class was recently observed in turbulent liquid crystals\cite{18,19}. 

In what concerns physical thermodynamic quantities, the order parameter distribution function has been proven 
to be an important tool for studying a large variety of subjects, as magnetic systems\cite{21,22,23,24,25},  
the liquid-gas critical point\cite{26}, the critical point in the unified theory of weak and electromagnetic 
interactions\cite{27}, and the critical point in quantum chromodynamics\cite{28}. 
To the best of our knowledge, even with all those applications, the method of the order parameter analysis did not receive much 
attention in nonequilibrium physics. In the present work, we explore the probability distribution 
of the order parameter of the GCP and the DKCA in order to analyze the DP universality class. Results show that the method is indeed quite reliable to study nonequilibrium phase transitions, and we hope that it could be generalized and applied to other dynamic models.

\section{\label{appr}Approach}

For the specific cases of the GCP and the DKCA, the order parameter can be 
chosen as the density of active sites, namely $\rho = \frac{1}{N} \sum_{i=1}^N \sigma_i$, where $N$ is 
the total number of sites and the occupation variable $\sigma_i$ is equal to $1$ ($0$) if site is active (inactive). 
In the infinite-size limit, $\rho$ vanishes in the absorbing phase. 
In finite-size systems, the density $\rho$ is a fluctuating quantity, 
characterized by the probability distribution $P(\rho)$. 
Analogous to the usual finite-size scaling assumptions\cite{29}, one then expects that, 
for a large finite system of linear dimension $L$ at the critical point, $P(\rho)$ takes the form
\begin{equation}
P(\rho) = bP^*(\tilde{\rho}), 
\label{Prho}
\end{equation}
where $b = b_0L^{\beta/\nu_{\perp}}$, $\beta$ and $\nu_{\perp}$ are the critical exponents of the density of 
active sites and the correlation length, respectively, $\tilde{\rho} = b\rho$, $b_0$ is a non-universal constant, and 
$P^*(\tilde{\rho})$ is a universal scaling function. For the DP universality class, one has $\beta = 0.276486(8)$ and 
$\nu_{\perp} = 1.096854(4)$ \cite{29a}. Scaling functions, such as that given by Eq. (\ref{Prho}), 
are characteristic of the corresponding universality class. Systems belonging to the same universality class 
share the same $P^*$ scaling function and thus, from the precise knowledge of $P^*(\tilde{\rho})$, 
one can characterize critical points and also identify universality classes.

The most efficient way to compute the probability distribution $P(\rho)$ 
is probably through Monte Carlo simulations. In equilibrium systems, $P(\rho)$ corresponds to the fraction 
of the total number of realizations in which the order parameter reaches the specific value $\rho$. 
In absorbing state systems, obtaining that distribution is a more complicated issue, since the active 
stationary distribution only appears in the infinite-size limit. For finite lattices, as the system always becomes 
trapped in the absorbing state, one can only evaluate the quasistationary (QS) distribution\cite{2}.

Let us briefly review the definition of the QS distribution. By denoting $n$ as the number of active sites 
($n=0$ corresponds to the absorbing state) and $P_n(t)$ as the probability of having exactly $n$ occupied 
sites at time $t$, the survival probability can be obtained by 
$p_s(t) = \sum_{n=1}^{N} P_n(t) = 1 - P_0(t)$. As $t \rightarrow \infty$, it is expected that $P_n$, normalized by the survival 
probability $p_s(t)$, remains time-independent\cite{2}. 
A procedure to compute the QS distribution is to restrict averages over the surviving realizations only, {\it i}. {\it e.}, 
after performing a large sample of independent realizations, the average value of some physical quantity at time {\it t} 
is taken over the realizations that did not reach the absorbing state at that time. At long times, as the number 
of surviving samples decays, this mechanism suffers from large fluctuations. 

A more effective way to compute the QS distribution was proposed by 
Tom\'e and de Oliveira\cite{30}. It consists in creating a particle in the finite system 
whenever the absorbing state is going to be reached. This procedure is equivalent to forbid the 
last particle to be annihilated and thus the density of active sites $\rho$ is always non-zero.
In the thermodynamic limit this perturbation was found to be irrelevant, as shown in Ref.\cite{30}.
The same authors have also proposed a conserved contact process in which $\rho$
is constant and the absorbing state is eliminated\cite{30a}. 
The model can be seen as 
the CP version in an ensemble of fixed particle number and its properties, in the thermodynamic limit,
are identical to those of the ordinary CP. The equivalence between both ensembles was shown by 
Hilhorst and van Wijland\cite{30b}.

Another powerful method to 
obtain QS distributions was proposed by de Oliveira and Dickman\cite{30c}. 
It consists in storing a list with $M$ non-absorbing 
configurations that the system has visited previously (typically $M \sim 10^3-10^4$). The list is updated 
with probability $p_{r}$ (usually $p_{r} \sim 10^{-3}-10^{-2}$), which means that a configuration from the list 
is replaced by the current configuration with probability $p_{r}$. During the simulation, if an absorbing 
configuration is imminent, it is replaced by another one, randomly chosen from the list. This procedure is 
used in the present work, with $M = 2000$ and $p_{r}=10^{-3}$ in most cases.

Regarding the simulation details, we have simulated the generalized contact process (GCP) in 
one-dimensional lattices with periodic boundary conditions and sizes $L$ varying 
from $80$ to $640$, and up to $3200$ in a few cases. Different starting configurations were tested, 
with an initial density of active sites $p_{i}$ varying from $0.2$ to $1$, and the QS distribution 
was found to be independent of $p_i$, within the error bars.  
For each lattice size, simulations of $10 - 1000$ samples with $10^6 - 10^7$ Monte Carlo 
steps per sample were performed.
Transition rates are schematically represented in Table~\ref{tab1}.
According to those transition rates, the time evolution can be described as following\cite{30d}.

\begin{itemize}
\item Choose a site $i$ randomly.
\item Choose a process (creation or annihilation):
      \begin{itemize}
      \item for $\zeta \leq 1$: choose creation with probability $\frac{1}{1+\mu}$ and 
                annihilation with probability $\frac{\mu}{1+\mu}$;  
      \item for $\zeta > 1$: choose creation with probability $\frac{\zeta}{\zeta+\mu}$ and 
                annihilation with probability $\frac{\mu}{\zeta+\mu}$.  
      \end{itemize}
\item If site $i$ is vacant ($\sigma_i=0$) and creation was chosen, one should define 
         $n = \sigma_{i-1} + \sigma_{i+1}$. Again, we need to consider both situations:
      \begin{itemize}
      \item for $\zeta \leq 1$: creation occurs with probabilities $0$, $\zeta$, and $1$ for 
               $n$ equal to 0, 1, and 2, respectively;
      \item for $\zeta > 1$: creation occurs with probabilities $0$, $1$, and $1/\zeta$ for 
               $n$ equal to 0, 1, and 2, respectively.
      \end{itemize}
\item After choosing $N$ sites, increase time by one unit.
\end{itemize}

\begin{table}[h]
\caption{Transition rates in the GCP. Open (filled) circles indicate vacant (occupied) sites.}
\begin{ruledtabular}
{\begin{tabular}{@{}ccccccc@{}} 
From & & &   To   & & & Rate \\ \colrule
$\circ    \circ \circ$    & &  &  $\circ    \bullet  \circ$     & &  &   0      \\
$\bullet \circ \circ$     & &  &  $\bullet \bullet  \circ$     & &  &  $\zeta$ \\
$\circ   \circ \bullet$   & &  &   $  \circ \bullet \bullet$   & &  &  $\zeta$ \\
$\bullet \circ \bullet$  & &  & $  \bullet \bullet \bullet$  & &  &   1 \\
$\bullet$                     & &  &  $\circ$                              & &  &  $\mu$ \\
\end{tabular} \label{tab1}}
\end{ruledtabular}
\end{table}

The one-dimensional DKCA was simulated on lattices with up to $3200$ sites and
averages were done over $10^3$ samples with $10^6$ Monte Carlo steps per sample. 
The transition probabilities 
$P[\sigma_i(t+1)|\sigma_{i-1}(t),\sigma_{i+1}(t)]$ were
$P[1|0,0]=0$, $P[1|0,1]=P[1|1,0]=p_1$, and $P[1|1,1]=p_2$. Naturally, 
$P[0|\sigma_{i-1},\sigma_{i+1}] = 1 - P[1|\sigma_{i-1},\sigma_{i+1}]$.
As already mentioned, in the DKCA all sites are updated simultaneously.

\section{\label{res}Results}

\begin{figure}[h]
\begin{centering}
\includegraphics[clip,angle=-90,scale=0.32]{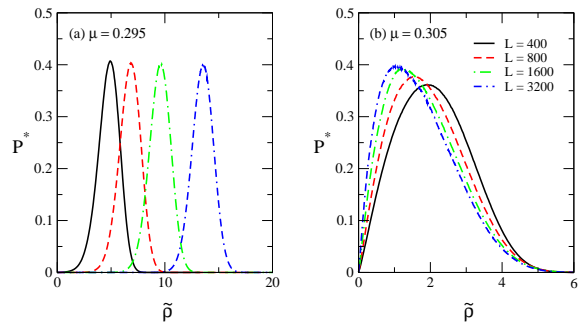}
\par\end{centering}
\caption{Probability distribution of the order parameter in the GCP for $\zeta = 0.5$. 
(a) $\mu = 0.295$ ($\mu < \mu_c$) and  (b) $\mu = 0.305$ ($\mu > \mu_c$). 
Error bars were omitted for better visualization and are typically around $0.3\%$.
\label{fig1}}
\end{figure}

\begin{figure}[htb]
\begin{centering}
\includegraphics[clip,angle=-90,scale=0.32]{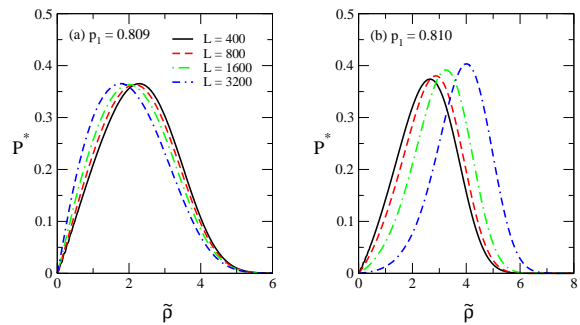}
\par\end{centering}
\caption{Probability distribution of the order parameter in the DKCA for $p_2 = 0$. 
(a) $p_1 = 0.809$  and  (b) $p_1 = 0.810$. 
Error bars were omitted for better visualization and are typically around $0.3\%$.
\label{fig2}}
\end{figure}

Following the mechanism proposed by Martins and Plascak\cite{24}, 
we analyzed the function $P^*(\tilde{\rho})$ to get an estimate of the critical point. 
As depicted in Fig.~\ref{fig1} one can see that, as the lattice size increases, 
the peak of the function moves to the right for $\mu = 0.295$, and it goes to the left 
for $\mu=0.305$. From a different point of view, let us consider the 
function $P^*(\tilde{\rho})$ for $L=400$ and $\mu_1=0.295$ as shown in Fig.~\ref{fig1}a.
The same distribution shall be obtained for a larger lattice (say for instance $L=800$) 
at a different rate $\mu_2$ in such a way that $\mu_2 > \mu_1$. On the other hand,
if we consider as reference the distribution for $L=400$ and $\mu_1=0.305$, 
we will have the same distribution for a larger lattice at $\mu_2 < \mu_1$. 
This suggests that the critical $\mu_c$ is in the range $0.295 < \mu_c < 0.305$.
The same behavior was observed for all other values of $\zeta$ as well as for the 
DKCA. Figure \ref{fig2} shows the normalized probability distribution function of the 
DKCA for $p_2=0$ and two values of $p_1$. 

\begin{figure}[htb]
\begin{centering}
\includegraphics[angle=-90,scale=0.31]{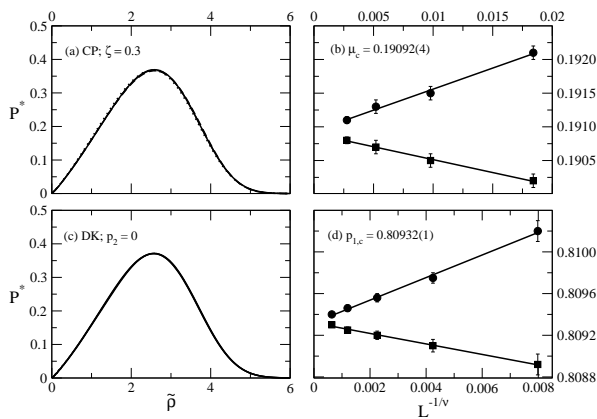}
\par\end{centering}
\caption{Graphs obtained from the data presented in Tabs. \ref{tab2}, \ref{tab3}, and \ref{tab4}. 
Error bars were omitted for better visualization. 
(a) Normalized probability distribution function for the GCP with $\zeta = 0.3$. The plot is, in fact, 
a superposition of four curves. 
(b) Corresponding finite-size scaling analyses for the GCP. 
(c) Normalized probability distribution function for the DKCA with $p_2 = 0$. The plot is, in fact, 
a superposition of five curves. 
(d) Corresponding finite-size scaling analyses for the DKCA.
\label{fig3}}
\end{figure}

In order to obtain a better estimate of the critical rate $\mu_c$ for the infinite lattice, 
one can proceed as following\cite{24}. 
By using a reference distribution function for a given $L$, $\zeta$, and $\mu$, 
one can vary $\mu$ for a different lattice size until a distribution that 
collapses into the reference one is obtained. For instance, 
Figure \ref{fig3}a shows the normalized probability distribution for $L=640$ at $\zeta=0.3$ and 
$\mu = 0.19080(3)$, considered as reference. For $L=320$, the same distribution is obtained at $\mu=0.19070(5)$. 
For $L=160$, the corresponding value of $\mu$ is 0.19050(5), while for $L=80$ one has 
$\mu=0.19020(10)$. All these four distributions are depicted in Fig. \ref{fig3}a.
Each one of those values of $\mu$ gives an estimate for the pseudo-critical $\mu_L$ for that lattice size. 
Since one expects that the difference $|\mu_L - \mu_c|$ scales as $L^{-1/\nu_{\perp}}$, where $\nu_{\perp}$ 
is the correlation length critical exponent, a finite-size scaling analysis can be performed 
to estimate the critical values of the infinite system. 
In Fig. \ref{fig3}b, one has a plot of $\mu_L$ vs. $L^{-1/\nu_{\perp}}$, 
with $\nu_{\perp} = 1.096854(4)$ (Ref. \cite{29a}). Each row in Table \ref{tab2} contains the values 
of $\mu_L$ that lead to the same distribution function for each $\zeta$ as well as the extrapolated 
value of $\mu_c$, obtained from the finite-size scaling technique. 
If another distribution is used as reference, a different set of $\mu_L$ is obtained 
(as shown in Table \ref{tab3}), providing another estimate for $\mu_c$. 
A similar analysis for the DKCA with $p_2=0$ is also depicted in Fig. \ref{fig3}, 
with the corresponding data represented in Tab. \ref{tab4}. A finite-size scaling analysis 
leads to a critical value $p_{1,c}=0.80932(1)$, which has even higher precision than previous works 
($p_{1,c} = 0.811(1)$ from \cite{31}, and $p_{1,c} = 0.8095(3)$ from \cite{32}).

\squeezetable
\begin{table}[htb]
\caption{Each row contains the annihilation rates $\mu_L$ at which the distribution functions are the same.  
Errors in parentheses affect the last digits. For $\zeta=0.01$ the lattice size $L=240$ was used instead of $L=80$, 
giving $\mu_L = 0.00834(1)$.
The last column shows the extrapolated values for $\mu_c(\zeta)$.}
\begin{ruledtabular}
{\begin{tabular}{@{}cccccc@{}} 
$\zeta$ &  L = 640   & L = 320      &   L = 160     &     L = 80    &  L $\rightarrow \infty$      \\  \colrule
0.01   & 0.00850(1)  & 0.00840(1)  & 0.00824(2)  &   $-$          &  0.00860(3) \\
0.02  & 0.01638(1)   & 0.01630(1)  & 0.01617(2)  & 0.01605(5) & 0.01641(3)  \\
0.05  & 0.03790(5)   & 0.0377(1)    & 0.0374(1)    & 0.0369(1)   & 0.03804(3)  \\
0.1    & 0.0709(1)    &  0.0706(1)    & 0.0701(1)    & 0.0694(2)   & 0.07111(7)    \\
0.2    & 0.1320(1)    &  0.1314(1)    & 0.1305(1)    & 0.1289(2)   & 0.13247(7)    \\
0.3    & 0.19080(3)  &  0.19070(5)  & 0.19050(5)  & 0.1902(1)   &  0.19090(2) \\
0.4    & 0.24740(2)  &  0.24720(3)  & 0.24685(5)  & 0.2465(1)   & 0.24750(8)  \\
0.5    & 0.3021(1)    &  0.3010(2)    & 0.2992(2)    & 0.2958(3)   &  0.3031(1)   \\
0.6    & 0.35660(5)  &  0.3552(1)    & 0.3528(3)    & 0.3488(5)   & 0.3578(2)    \\
1.0    & 0.5735(1)    &  0.5728(2)    & 0.5718(3)    & 0.5700(5)   &  0.5740(1)   \\
2.0    & 1.1030(1)    &  1.1023(1)    & 1.1015(3)    & 1.1001(8)   &  1.1034(1)   \\
\end{tabular} \label{tab2}}
\end{ruledtabular}
\end{table}

\squeezetable
\begin{table}[htb]
\caption{Sets of $\mu_L$ obtained from another reference distribution. 
Again, for $\zeta=0.01$, $L=240$ was used instead of $L=80$, 
giving $\mu_L = 0.00876(1)$.
}
\begin{ruledtabular}
{\begin{tabular}{@{}cccccc@{}} 
$\zeta$ &  L = 640   & L = 320      &   L = 160     &     L = 80      &  L $\rightarrow \infty$      \\  \colrule
0.01   & 0.00870(1)  & 0.00873(1)  & 0.00882(2)  &       $-$        &  0.00865(3) \\
0.02  & 0.01655(1)   & 0.01659(2)  & 0.01672(4)  & 0.01705(10) & 0.01643(3)  \\
0.05  & 0.03820(1)   & 0.03827(3)  & 0.03840(6)  & 0.03870(10) & 0.03810(3)  \\
0.1    & 0.07150(2)  &  0.07175(6)  & 0.0722(1)    & 0.0730(1)     & 0.07125(8)    \\
0.2    & 0.13300(5)  &  0.13330(5)  & 0.1339(1)    & 0.1350(2)     & 0.13264(7)    \\
0.3    & 0.19110(5)  &  0.1913(1)    & 0.1915(1)    & 0.1921(1)     &  0.19094(4) \\
0.4    & 0.24780(5)  &  0.24800(5)  & 0.2483(1)    & 0.2492(2)     & 0.24752(7)  \\
0.5    & 0.3040(1)    &  0.3046(3)    & 0.3058(4)    & 0.3080(5)     &  0.3033(1)   \\
0.6    & 0.3585(1)    &  0.3588(1)    & 0.3592(1)    & 0.3607(4)     & 0.3580(2)    \\
1.0    & 0.5750(1)    &  0.5757(1)    & 0.5770(3)    & 0.5800(5)     &  0.5740(1)   \\
2.0    & 1.1042(1)    &  1.1045(1)    & 1.1053(3)    & 1.1072(8)     &  1.1035(1)   \\
\end{tabular} \label{tab3}}
\end{ruledtabular}
\end{table}

\begin{table*}[htb]
\caption{Results for the DKCA with $p_2=0$. Each row shows the values of $p_{1,L}$ at which the normalized 
probability distribution is the same. These data were used to plot Fig. \ref{fig3}d and give the 
estimate of the critical $p_{1,c}$ in the last column.
}
\begin{ruledtabular}
{\begin{tabular}{@{}ccccccc@{}} 
$L$          & 3200           & 1600          & 800            & 400            & 200              & $L \rightarrow \infty$ \\ \colrule 
$p_{1,L}$ & 0.80940(1)  & 0.80946(3) & 0.80956(4) & 0.80975(5) & 0.8102(1)     & 0.80932(1)  \\
$p_{1,L}$ & 0.80930(1)  & 0.80925(3) & 0.80920(4) & 0.80910(6) & 0.80892(10) & 0.80932(1)  \\
\end{tabular} \label{tab4}}
\end{ruledtabular}
\end{table*}

\begin{figure}[htb]
\begin{centering}
\includegraphics[clip,angle=-90,scale=0.32]{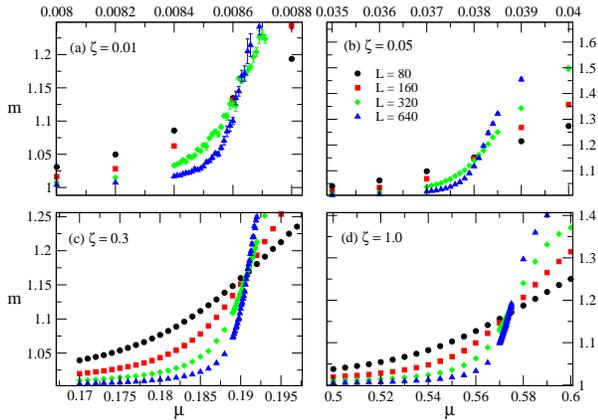}
\par\end{centering}
\caption{Moment ratio $m$ (as defined in text) as a function of the annihilation rate $\mu$.
Error bars when not shown are smaller than the symbols.
\label{fig4}}
\end{figure}

An often used technique to study the criticality of the DP universality class
consists in evaluating the moment ratio $m = \langle \rho^2 \rangle / \langle \rho \rangle^2$\cite{33}.
This quantity is analogous to the reduced fourth cumulant\cite{21} and reaches a universal 
value at the critical point. Thus, the curves for $m(\mu,L)$ cross near $\mu_c$ for different $L$.
Figure \ref{fig4} illustrates $m(\mu,L)$ for different values of $\zeta$.
To compare the results obtained by using the probability distribution function to those 
coming from the crossings of the moment ratio, Table \ref{tab5} shows the critical $\mu_c$ 
achieved by both methods. The results that come from the probability distribution are 
the mean value of the extrapolated $\mu_c$ depicted in Tables \ref{tab2} and \ref{tab3} for the 
generalized contact process and in Table \ref{tab4} for the Domany-Kinzel cellular automaton.

\begin{table}[htb]
\caption{Comparison between the critical values from two different methods.}
\begin{ruledtabular}
{\begin{tabular}{@{}ccc@{}} 
$\zeta$ & $\mu_c$ from         & $\mu_c$ from      \\ 
             & cumulant crossings & probability distributions \\ \colrule
0.01   & 0.00865(3)   & 0.00863(3)  \\
0.02   & 0.01648(3)   & 0.01642(2)  \\
0.05   & 0.03810(6)   & 0.03807(4)  \\
0.1     & 0.0713(1)     & 0.07118(8)  \\
0.2     & 0.1326(1)     & 0.13256(8)  \\
0.3     & 0.1909(1)     & 0.19092(4)  \\
0.4     & 0.2476(1)     & 0.24751(8)  \\
0.5     & 0.3032(1)     & 0.3032(1)    \\
0.6     & 0.3582(2)     & 0.3579(2)    \\
1.0     & 0.5742(3)     & 0.5740(1)    \\
2.0     & 1.1038(5)     & 1.10345(15) \\ 
DKCA & 0.8093(1)     & 0.80932(1)   \\
\end{tabular} \label{tab5}}
\end{ruledtabular}
\end{table}

\section{\label{conc}Conclusions}

In focusing on the study of the probability distribution of the order parameter in systems that do not obey 
the detailed balance, this work considered two different models in the directed percolation universality class.
The generalized contact process and the Domany-Kinzel cellular automaton were investigated.
The criticality of both models was obtained by using the probability distribution of the order parameter itself
and the results showed that this approach is also powerful to study nonequilibrium phase transitions, 
regarding their universal and nonuniversal aspects. 
In general, the critical values obtained from the present method have higher precision 
than the values from the crossings of the moment ratio. 
In addition, this work has provided an accurate estimate for the critical point in the Domany-Kinzel cellular automaton 
with $p_2=0$. To the best of our knowledge, the present approach, using just the probability distribution 
of the order parameter as expressed in Eq. (\ref{Prho}), was applied to nonequilibrium systems for the first time. 
We believe that these results will spread the treatment of other dynamic systems within the present approach. 
Applications to damage spreading transitions, that are supposed to be in the same universality class, are now in progress.

\section*{Acknowledgments}
This work was supported by the Coordena\c c\~ao de Aperfei\c coamento de Pessoal de N\'{\i}vel Superior (CAPES), 
Brazil (Grant No. 1144-10-3) and Funda\c c\~ao de Amparo \`a Pesquisa do Estado de Mato Grosso (FAPEMAT), 
Brazil (Grant No. 461884/2009). 
Author is indebted to D. P. Landau for a critical reading of the manuscript and to 
R. Dickman and E. Mello Silva for interesting discussions.

\bibliography{references.bib}

\end{document}